\documentclass[aps,pra,twocolumn,amsmath,amssymb,a4paper,superscriptaddress]{revtex4}
\usepackage{graphicx} 
\usepackage{epstopdf}
\usepackage{color}
\usepackage{hyperref}

\hypersetup{%
   pdfpagemode=None, 
   pdfstartpage=1,
   pdfstartview=FitH,
   pdfmenubar=true,
   pdftoolbar=true,
   colorlinks = true,
   linkcolor=blue,
   citecolor=blue,
   bookmarksopen=false
 }

\begin{document}

\title{Enhancing photoassociation rates by non-resonant light control
  of shape resonances} 

\author{Rosario Gonz\'alez-F\'erez}
\affiliation{Instituto ‘Carlos I’ de F\'isica
    Te\'orica y Computacional and Departamento de F\'isica At\'omica,
    Molecular y Nuclear, Universidad de Granada, 18071 Granada,
    Spain}

\author{Christiane P. Koch}
\affiliation{Theoretische Physik, Universit\"at Kassel,
  Heinrich-Plett-Str. 40, 34132 Kassel, Germany} 
\date{\today}

\begin{abstract}
  Photoassociation, assembling molecules from atoms using laser light,
  is limited by the low density of atom pairs at sufficiently short
  interatomic separations. Here, we show that non-resonant light with
  intensities of the order of $10^{10}\,$W/cm$^2$  modifies the
  thermal cloud of atoms, enhancing the Boltzmann weight of shape
  resonances and pushing scattering states below the dissociation
  limit. This leads to an enhancement of photoassociation rates  by
  several orders of magnitude and opens the way to significantly
  larger numbers of ground state molecules in a thermal ensemble than
  achieved so far.  
\end{abstract}
\maketitle

\section{Introduction}
\label{sec:intro}

The formation of ultracold molecules ($T\le 100\,\mu$K) has
experienced rapid 
progress over recent years~\cite{ColdMolBook}. At ultralow
temperatures, extremely precise control over the molecules' dynamics
can be exerted. This allows for transferring  molecules into a
single quantum state and even for controlling their
reactivity~\cite{OspelkausSci10}. Their rich internal structure makes them 
sensitive probes in precision measurements of fundamental
constants~\cite{ZelevinskyPRL08}. Possibly strong 
dipolar interactions facilitate their use in quantum computation and
quantum simulation~\cite{MicheliNatPhys06}. 

To produce molecular samples at ultralow temperatures, molecules
are typically assembled from ultracold atoms using
magnetic-field controlled Feshbach resonances or laser light.
As demonstrated in a series of spectacular experiments, 
ultracold molecules in their internal ground state are obtained 
by subsequently applying Stimulated Raman Adiabatic Passage
(STIRAP)~\cite{LangPRL08,NiSci08,DanzlSci08}   
or broadband vibrational cooling~\cite{PilletSci08}. However, in all
these experiments, the number of ground state molecules that can be
produced has so far been limited to about $10^{4}$. 
When the molecules are formed by magnetoassociation, this small number is
explained by the very low temperature, of the
order of hundreds of nano-Kelvin, and corresponding high phase
space density, attained by e.g. evaporative cooling. In the
case of photoassociation, usually carried out in magneto-optical
traps holding up to 10$^{10}$ atoms at temperatures between
$1-100\,\mu$K and a phase space density many orders of magnitude
lower, only a small fraction of atom pairs resides at  
sufficiently short interatomic separations to be photoassociated. 
The number of ground state molecules produced directly by photoassociation has thus
also been limited to about $10^{4}$~\cite{DeiglmayrPRL08}.
A significant increase of the photoassociation rate is the problem
that we address here.  

To this end, we suggest to control shape resonances using non-resonant
light to enhance photoassociation rates. Non-resonant light
couples to the anisotropic polarizability of an atom pair, shifting
the position of shape resonances to lower energies~\cite{shaperes}
and increasing the resonance's
thermal weight in an ultracold trap. Below we demonstrate an
enhancement by three orders of magnitude in the photoassociation rate of
strontium, a molecule which is currently attracting considerable
attention~\cite{Schreck12,Zelevinsky12,SkomorowskiKochJCP12,SkomorowskiKochPRA12}. 
Photoassociation relies only on the presence of optical transitions
which usually are abundant. 
Shape resonances are ubiquitous for diatomics. The
light-matter coupling via the anisotropic polarizability is of
universal character, independent of a particular energy level
structure, frequency of the light, or presence of a permanent dipole
moment. Therefore our findings are easily extended to 
molecules other than Sr$_2$. 

\begin{figure}[tb]
  \centering
  \includegraphics[width=0.95\linewidth]{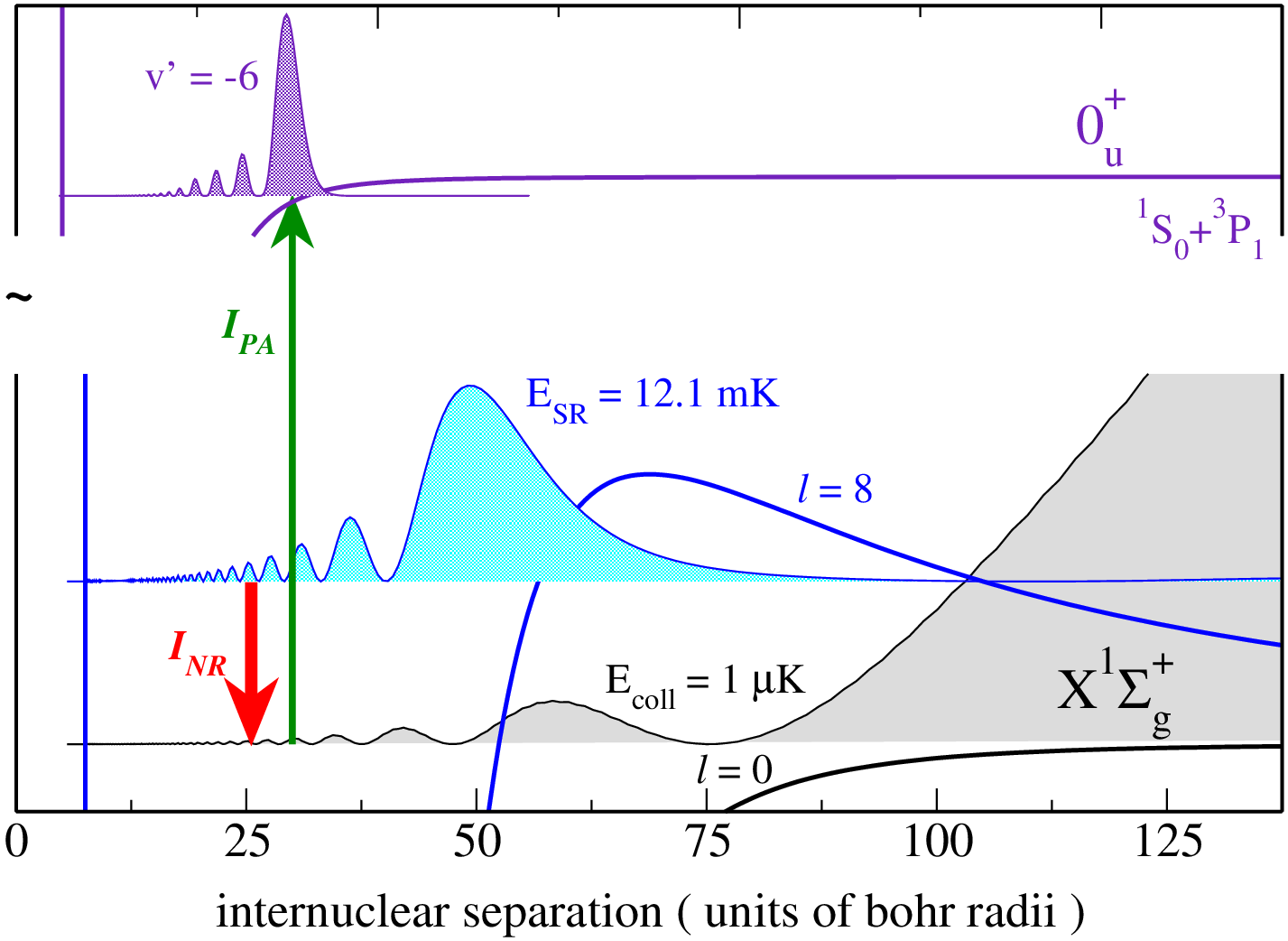}
  \caption{(Color online)
    left: The photoassociation efficiency is determined by the
    free-bound Franck-Condon overlap between the scattering states and
    weakly bound excited state levels (purple line), shown
    here for photoassociation of $^{88}$Sr$_2$ molecules near an intercombination
    line transition~\cite{ZelevinskyPRL06,Schreck12,Zelevinsky12}.  
    For $s$-waves (grey
    line, scaled by a factor of $10^6$ compared to the blue
    line), the probability density at short internuclear
    distances is very small, leading to low photoassociation efficiency. A shape
    resonance (blue line corresponding to a field-free rotational quantum
    number $l=8$) is a quasi-bound state trapped
    behind the rotational barrier. Typically, its energy, $E_{SR}$, is too high
    for the shape resonance to carry any significant thermal weight in
    an ultracold cloud.
    Here we suggest to employ a strong non-resonant field (red arrow)
    which leads to rotational  
    hybridization of the scattering states, effectively moving the
    position of the shape resonance to lower collision energies, $E_{coll}$, and
    increasing the thermal weight of the shape resonance.
  }
  \label{fig:scheme}
\end{figure}

Our approach is related to Feshbach-optimized
photoassociation~\cite{PellegriniPRL08} which also employs
a scattering resonance to increase the number of atom
pairs that are quasi-trapped at short interatomic separations. 
Instead of tuning an existing resonance, one could also use resonances
induced by an external field, for example, electric-field induced
resonances for polar
molecules~\cite{KremsPRL06,RosarioNJP09}. However, the electric fields
required to achieve a significant enhancement (of 
about 1.3$\,$MV/cm in the most favorable case) are
currently experimentally unfeasible~\cite{ChakrabortyJPB11}.
Unlike Feshbach-optimized photoassociation, shape resonance control 
does not require the presence of a hyperfine manifold of the atoms. 
Shape resonances occur naturally for partial waves with
$l>0$ when a scattering state becomes trapped behind the centrifugal
barrier. Since shape resonances are quasi-bound states, they have very
favorable free-to-bound overlaps required for efficient photoassociation;
cf. the blue line in Fig.~\ref{fig:scheme} (left). 
In some fortitious cases such as
Rb$_2$~\cite{BoestenPRL96,BoestenPRA97} or Cs$_2$, the energy of a
shape resonance almost matches the trap temperature, 
and large photoassociation rates are observed. Typically, however, the lowest
energies at which shape resonances occur are far from the trap  
temperature, and their thermal weight is  very small. In this case, 
photoassociation is limited by the atom pair density $\rho(R)$ at interatomic
separations $R \lesssim 200\,$a$_0$~\cite{KochJPhysB06}; i.e., it 
addresses mostly $s$-wave atom pairs for which free-to-bound
Franck-Condon factors are very small, cf. the grey line in 
Fig.~\ref{fig:scheme} (left). This can be changed by applying a
strong non-resonant field prior to and during photoassociation. 

\section{Theoretical framework}
\label{sec:theo}

We assume the non-resonant light, with intensity $I$, and a photoassociation laser, 
with intensity $I_{PA}$, to be linearly polarized along the $z$-axis
of the laboratory fixed frame. Then, in the body fixed frame, the
interaction of a pair of atoms with these two fields is described by
the Hamiltonian 
\begin{equation}
  \label{eq:PA-Hamil}
   H =
  \begin{pmatrix}
     H_g & 
    -\frac{D( R)}{2}\sqrt{\frac{I_{PA}}{2\epsilon_o c}} \cos\theta \\
    -\frac{D( R)}{2}\sqrt{\frac{I_{PA}}{2\epsilon_o c}} \cos\theta &
     H_e + \Delta_{PA}  
  \end{pmatrix} 
\end{equation}
with
\begin{eqnarray*}
  \label{eq:Hi}
   H_j =  T + V_j( R) + \frac{\vec{L}^2}{2\mu R^2}
  -\frac{I}{2\epsilon_0 c} \left[\Delta\alpha^j( R)\cos^2\theta 
    + \alpha^j_{\perp}( R)\right] 
\end{eqnarray*}
and $j=g,e$ for the electronic ground and excited state,
respectively. 
$\Delta_{PA} = \hbar(\omega_{PA}-\omega_{at})$ denotes
the detuning of the photoassociation laser from the atomic resonance, 
$\epsilon_0$ and $c$ are the electric vacuum permittivity and speed of
light. The photoassociation laser interacts with the transition dipole
moment, $D( R)$, coupling the electronic ground state to an
electronically excited state. $V_g( R)$ and $V_e( R)$ denote the
respective potential energy curves and $ T_R$ and $\vec{L}^2/2\mu
R^2$ are the vibrational and rotational kinetic energies. 
The non-resonant light intensity couples to the polarizability
anisotropy, $\Delta\alpha^j( R)=\alpha^j_{\parallel}(
R)-\alpha^j_{\perp}( R)$ ($j=g,e$), where 
$\alpha^j_{\parallel/\perp}( R)$ are the parallel and perpendicular
molecular polarizability components. 
The behaviour at large internuclear distances is well approximated
by Silberstein's formula in terms of atomic
polarizabilities~\cite{HeijmenMP96,JensenJCP02}.
We consider $^{88}$Sr atoms trapped at a temperature of 1$\,\mu$K. 
The potential energy curve for the electronic ground state was taken from
Ref.~\cite{SteinPRA08} and adjusted to yield a scattering length of
$a_S=-2\,a_0$~\footnote{The photoassociation results are not affected when varying
  the scattering lengths from $a_S=+4.5\,a_0$ to $a_S=-4.5\,a_0$.}.
The potential energy curves for the electronically excited
A$^1\Sigma_u^+$, c$^3\Pi_u$, and a$^3\Sigma_u^+$ manifold, spin-orbit
coupling functions and transition dipole matrix elements are found in
Ref.~\cite{SkomorowskiKochJCP12}. We investigate photoassociation into the lowest
level below the $^1S+^3P$ asymptote that was previously observed
($v'=-6$)~\cite{ZelevinskyPRL06}. This level is well represented
by an adiabatic potential energy curve $0_u^+$ neglecting resonant
spin-orbit coupling~\cite{SkomorowskiKochPRA12}. Its spontaneous
emission lifetime was calculated to be
$\gamma_{v'=-6}=7.6\,\mu$s~\cite{SkomorowskiKochPRA12}. The atomic
polarizabilities for ground and excited state are taken from
Refs.~\cite{CRChandbook,PorsevPRA08}.  
The Hamiltonian~\eqref{eq:PA-Hamil} is represented by a 
mapped Fourier grid for the radial part and a basis set expansion 
in terms of Legendre polynomials for the angular part,
taking advantage of the magnetic quantum number $m$ being conserved. 
About $N_R\approx 2000$
radial grid points and $l_{max}=20$ are required to converge the
calculations for $T=1\,\mu$K and $I\le 1.5 \times
10^{10}$W/cm$^2$. 

Diagonalizing the Hamiltonian \eqref{eq:PA-Hamil} for $I_{PA}=0$ 
yields the bound levels and continuum states of the ground and excited
electronic states in the presence of the non-resonant field. 
For $I \neq 0$, only the magnetic quantum number is conserved. The
non-resonant field mixes different partial waves of the same parity
and $l$ and $l'$ are not good quantum number
any more. For the sake of simplicity, we will label the field-dressed
states by the field-free quantum numbers $l,m$, and $v',l',m'$, adding
a tilde to $l$, $v'$ and $l'$ to indicate that they are labels not quantum
numbers. Note that for
bound states, the field-dressed levels $\tilde{v}'$, $\tilde{l'}$ are
adiabatically connected to the 
field-free quantum numbers $v'$, $l'$ even for very large intensities. 
The hybridization of the angular motion of
the field-dressed wave functions is analyzed in terms of 
the rotational weights of the field-free partial
waves,
\begin{equation}
  \label{eq:c_l}
  c_l = \int dE \int dR \int d \cos\theta \,\psi^g_{\tilde{l},0}(R,\theta;E_{SR})
  \psi^g_{l,0}(R,\theta;E)\,,  
\end{equation}
where $\psi^g_{\tilde{l},0}(R,\theta;E_{SR})=
\langle R,\theta|\psi_{\tilde{l}m=0}^g(E_{SR})\rangle$, and 
$\psi^g_{l,0}(R,\theta;E)$ are the field-free scattering wavefunctions for
partial wave $l$. 
The photoassociation rate coefficient in the presence of the non-resonant field is
determined from the excited state bound levels and ground state
continuum states, $|\psi^e_{\tilde{v}'\tilde{l'}m'}\rangle$ and
$|\psi_{\tilde{l}m}^g(E)\rangle$ with 
$|\psi_{\tilde{l}m}^g(E)\rangle$ normalized with respect to
energy~\cite{RosarioPRA07}.  

For an ensemble of atom pairs with a Maxwell-Boltzmann velocity
distribution, employing an isolated photoassociation resonance, the photoassociation rate
coeffcient is given by~\cite{NapolitanoPRL94}  
\begin{widetext}
\begin{equation}
  \label{eq:Kgeneral}
  K_{\tilde{v}'}(I_{PA},\omega_{PA},I,T) = \frac{k_BT}{h Q_T} \int_0^\infty 
  \sum_{\tilde{l}=0}^\infty\sum_{\tilde{l'}=0}^\infty  e^{-\frac{E}{k_BT}}
  \frac{\gamma_{\tilde{v}'\tilde{l}'m'}\gamma_s(I_{PA},\tilde{v}',\tilde{l}',m',E,\tilde{l},m,I)}
  {(E-\Delta_{\tilde{v}'\tilde{l}'m'})^2+(\gamma/2)^2}
\frac{dE}{k_B T}
\end{equation}  
\end{widetext}
with $\gamma_{\tilde{v}'\tilde{l}'m'}$ the spontaneous decay rate of level
$\tilde{v}'\tilde{l}'m'$, approximated by $\gamma_{v'=-6,l'=0,m'=0}$, 
$\gamma_s\equiv\gamma_s(I_{PA},\tilde{v}',\tilde{l}',m',E,\tilde{l},m,I)$
the stimulated emission 
rate, and $\gamma$ the total width, $\gamma=\gamma_{\tilde{v}'\tilde{l}'m'}+\gamma_s$. 
$Q_T=(2\pi \mu k_BT/h^2)^{3/2}$ denotes the translational
partition function, and $\Delta_{\tilde{v}'\tilde{l}'m'}$
the detuning, $\Delta_{\tilde{v}'\tilde{l}'m'}=E_{\tilde{v}'\tilde{l}'m'}-\hbar\omega_{PA}$. 
Since $^{88}$Sr$_2$ is a bosonic molecule with zero nuclear spin, the
sum in Eq.~\eqref{eq:Kgeneral} runs solely over even partial waves
$\tilde{l}$ in the electronic ground state.
We consider only $m=m'=0$ in Eq.~\eqref{eq:Kgeneral}, for
which the effect of the non-resonant field is largest. The actual
photoassociation rate will be slightly higher than predicted here due to the
neglected contributions from terms with $m \neq 0$ to the sum of
Eq.~\eqref{eq:Kgeneral}. 
For weak photoassociation intensities, the stimulated emission rate
can be approximated by Fermi's golden rule, 
\begin{eqnarray*}
  &&\gamma_s(I_{PA},\tilde{v}',\tilde{l}',m',E,\tilde{l},m,I) \approx\\
  &&\quad\quad\quad\frac{4\pi^2I_{PA}}{c} |\langle\psi^e_{\tilde{v}'\tilde{l}'m'}| D( R)\cos 
  \theta|\psi_{\tilde{l}m}^g(E)\rangle|^2 \,,
\end{eqnarray*}
i.e., by an integral over $R$ and $\theta$, ensuring the correct
selection rules. 

\section{Results: Photoassociation of Sr$_2$}
\label{sec:Sr2}

\begin{figure}[tb]
  \centering
  \includegraphics[width=0.95\linewidth]{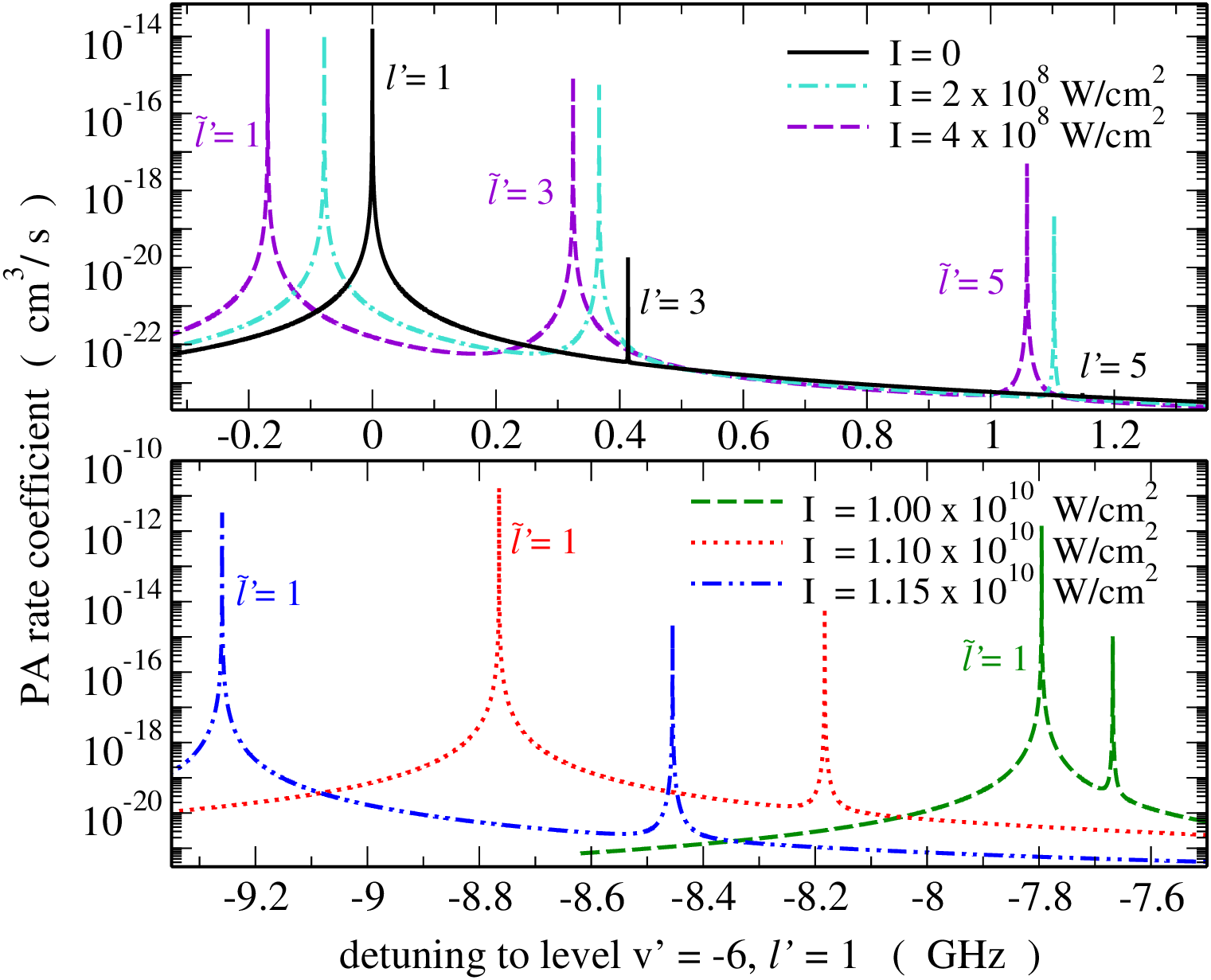}
  \caption{(Color online)
    photoassociation lines (absorption rate coefficient) for photoassociation into
    $\tilde{v}'=-6$ for weak (top) and strong (bottom) non-resonant fields
    $I$. The intensity of the photoassociation laser is set to
    $I_{PA}=1\,$W/cm$^{2}$, and the trap temperature to 
    $T=1\,\mu$K. 
  }
  \label{fig:PA_Sr2}
\end{figure}

The photoassociation rate, $K_{\tilde{v}'}(I_{PA},\omega_{PA},I,T)$, 
is shown in Fig.~\ref{fig:PA_Sr2} for $\tilde{v}'=-6$ and 
weak (top) and strong (bottom) non-resonant fields.  
The black solid line shows the typical rotational progression of a photoassociation
spectrum for $I=0$. The largest peak is observed for $l'=1$,
reflecting the pure $s$-wave character of the thermal cloud at
$1\,\mu$K. A weak non-resonant field (Fig.~\ref{fig:PA_Sr2}, top) shifts
the photoassociation peaks. The largest shift occurs for $\tilde{l}'=1$.
For the optical lattice of Ref.~\cite{Zelevinsky12} with $I \sim
10^{7}$W/cm$^2$, the shift amounts to about 3 MHz. An increase of the peak
heights is only observed for $\tilde{l}'=3$ and $\tilde{l}'=5$. It is
rationalized in terms of rotational hybridization in the excited state
levels. Generally, bound levels are more strongly affected by the
non-resonant field than scattering states. The $\tilde{l}'=3$ and
$\tilde{l}'=5$ states have thus some 
contribution from the $l'=1$ partial wave which renders photoassociation from
$s$-wave atom pairs into the field-dressed $\tilde{l}'=3$ and
$\tilde{l}'=5$ bound state allowed, enhancing the  corresponding  photoassociation
peak.   
While this increase amounts to several orders of magnitude, the total
photoassociation rate for $\tilde{l}'=3$ and $\tilde{l}'=5$ is still smaller than
the $l'=1$ photoassociation rate without non-resonant field. 

\begin{figure}[tb]
  \centering
  \includegraphics[width=0.9\linewidth]{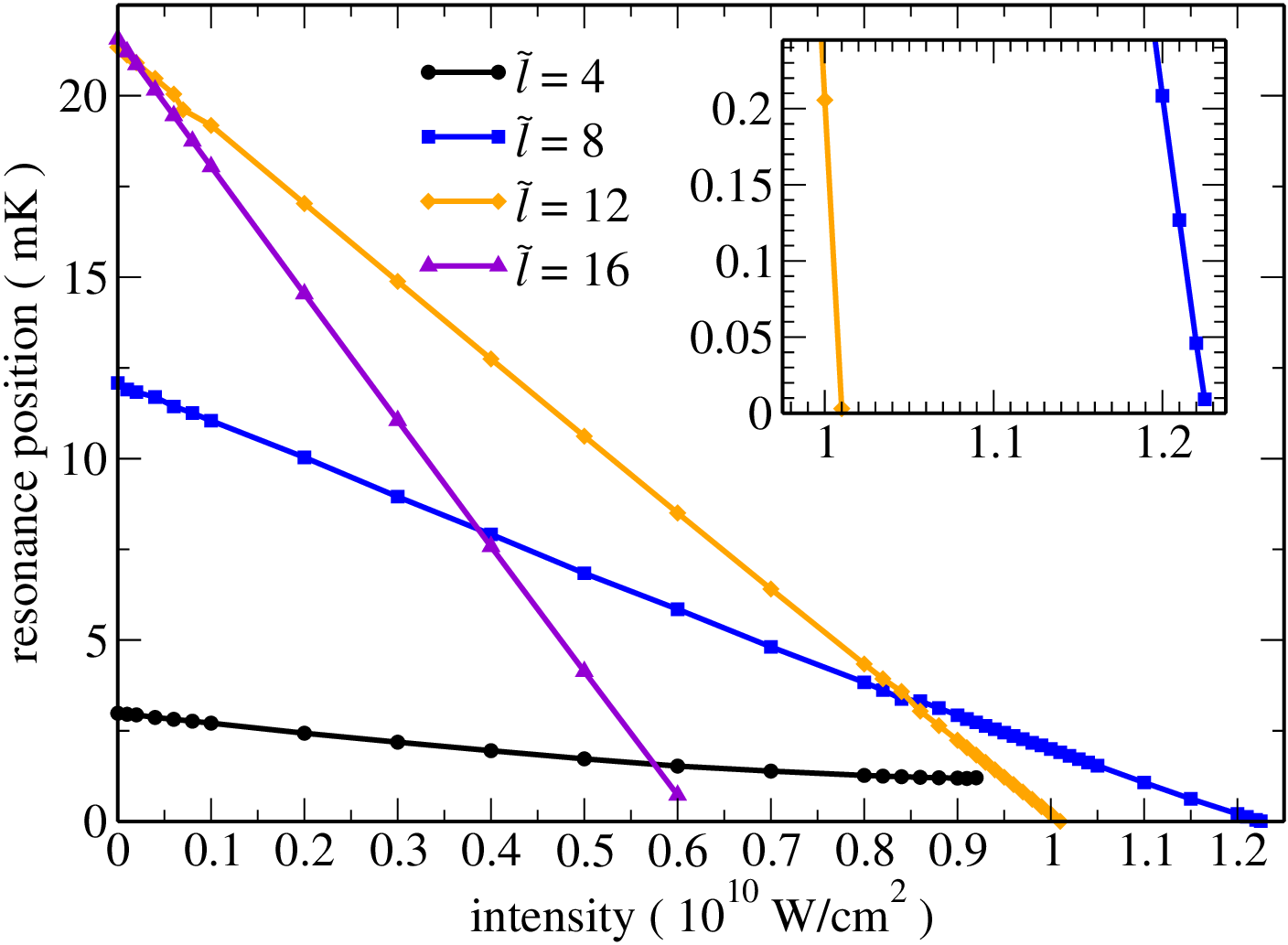}  
  \caption{
    Energy of the $\tilde{l}=4$,
    $\tilde{l}=8$, $\tilde{l}=12$ and $\tilde{l}=16$ shape resonances
    vs non-resonant field intensity.}
  \label{fig:respos}
\end{figure}
\begin{figure}[bt]
  \centering
  \includegraphics[width=0.99\linewidth]{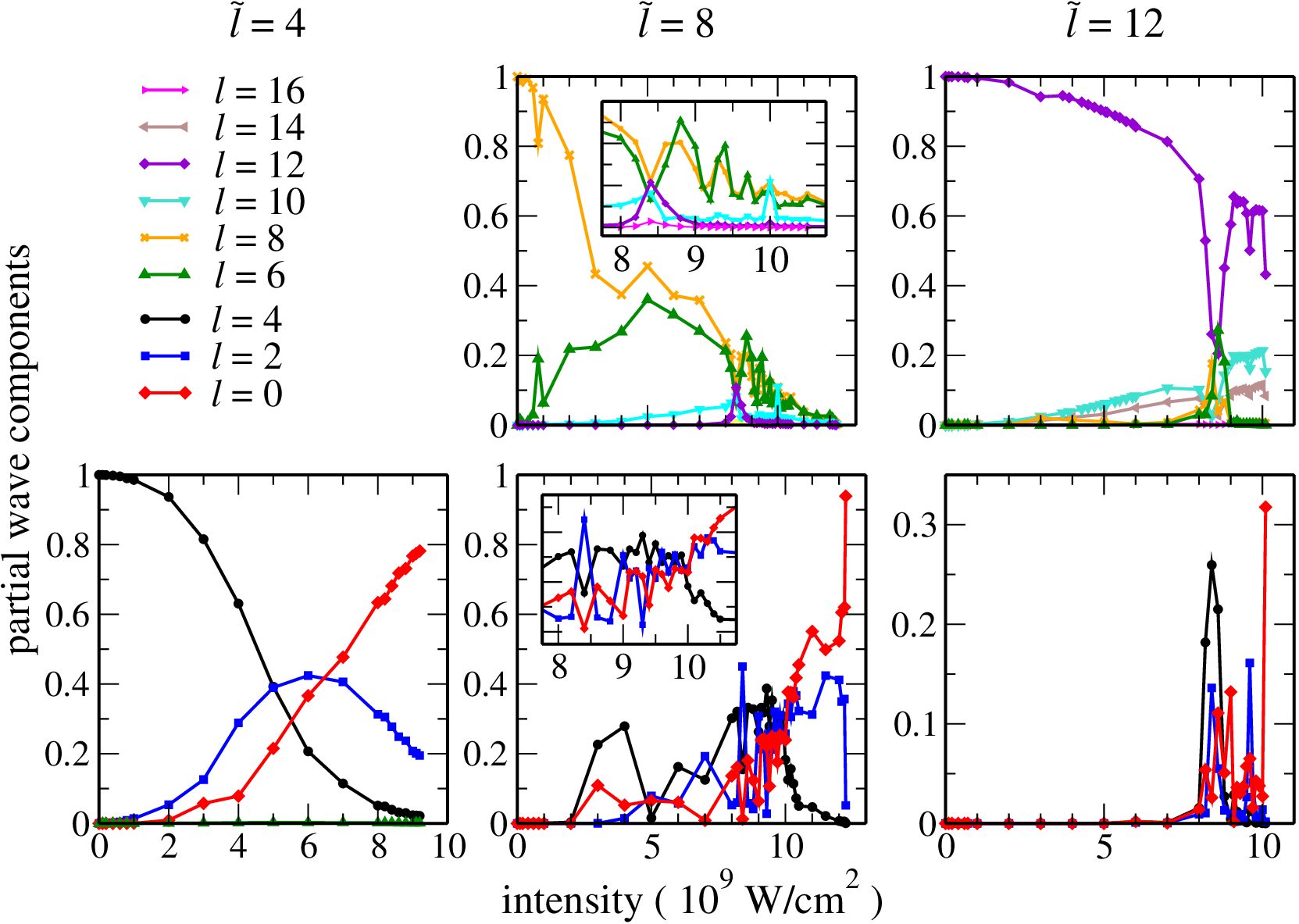}
  \caption{(Color online)
    Contribution of different partial waves $l$ to the resonance
    wavefunction for the $\tilde{l}=4,8,12$ shape resonances as a function 
    of the non-resonant intensity. The low-$l$ contributions are shown
    in the lower panel, the high-$l$ contributions in the upper
    panel. For high non-resonant field intensities, all resonance
    wavefunctions acquire predominantly $s$-wave character.
  }
  \label{fig:rotcomp}
\end{figure}
At high non-resonant light intensities 
(Fig.~\ref{fig:PA_Sr2}, bottom), the rotational progression observed
for $I=0$ is replaced by the pendular progression. More 
remarkably, a large increase of the $\tilde{l}'=1$ peak, amounting to
about three orders of magnitude compared to the field-free $l'=1$ photoassociation
rate, is observed in Fig.~\ref{fig:PA_Sr2}  for non-resonant field
intensities of the order of $10^{10}$W/cm$^2$. For  $T=1\,\mu$K,
the $\tilde l=12$ shape resonance  acquires its highest thermal weight 
close to $I\sim 10^{10}$W/cm$^2$, cf. Fig.~\ref{fig:respos} showing
the resonance position as a function of non-resonant field intensity.
The remarkable enhancement 
of the $\tilde{l}'=1$ photoassociation rate in Fig.~\ref{fig:PA_Sr2}
is thus related to the field-dressed
$\tilde{l}=12$ shape resonance. This is further analyzed by inspection
of the rotational weights, Eq.~\eqref{eq:c_l}, of the 
field-dressed resonance wavefunction shown in
Fig.~\ref{fig:rotcomp}. 
The lower (upper) panel of Fig.~\ref{fig:rotcomp} displays the
components for low (high) $l$. 
Note that the coupling mixes only partial waves of the same
parity. For $I=0$, the resonance wavefunctions are pure
$l=4$, $l=8$ and $l=12$ states, respectively. 
As the non-resonant field is switched on, a substantial
amount of first $l-2$ and then $l-4$ is mixed in to the field-free
states. For large non-resonant field intensities, the trend of mixing
in lower-$l$ components is observed all the way down to $l=0$ (red
diamonds).Thus, photoassociation from the field-dressed
resonance wave function into $\tilde{l}'=1$ becomes
possible for large $I$, explaining the large
enhancement of photoassociation 
into an excited state level with $\tilde l'=1$ observed in
the lower panel of Fig.~\ref{fig:PA_Sr2}. 

While the dependence on the non-resonant intensity in
Fig.~\ref{fig:rotcomp} is smooth for the $\tilde{l}=4$ resonance,
discontinuous  
behavior is observed for $\tilde{l}=8$, $\tilde{l}=12$. Each of the
features in the apparantly complicated dependence for  $\tilde{l}=8$,
$\tilde{l}=12$ can be rationalized in terms of non-adiabatic behavior
of interacting resonances: The first kink in the dependence of the
$\tilde{l}=8$ resonance is observed at $8\times 10^8\,$W/cm$^2$. It is
due to the interaction with a $\tilde l=6$ shape resonance
~\footnote{The $\tilde l=6$ shape resonance can be
  identified 
  by inspection of the scattering wavefunctions. However, it is too
  diffuse for its energy to be determined with sufficient precision
  and was therefore omitted from Fig.~\ref{fig:respos}.}.
A second discontinuous feature for $\tilde l =8$ is observed around
$4\times 10^9\,$W/cm$^2$. It is caused by the interaction of the $\tilde
l=8$ and $\tilde l=16$ resonances close to their crossing, see
Fig.~\ref{fig:respos}. The most striking non-adiabatic features occur for 
both $\tilde{l}=8$ and $\tilde{l}=12$ resonances between $8\times
10^9\,$W/cm$^2$ and $1\times 10^{10}\,$W/cm$^2$. It is due to the
non-adiabatic interaction between these two resonances which cross near
$8.5\times 10^{9}\,$W/cm$^2$ and with other scattering states, 
introducing a strong mixing of the resonance wavefunctions.
The non-adiabatic behavior of the shape resonances observed in
Fig.~\ref{fig:rotcomp} underlines the requirement to treat
the fully coupled rotational-vibrational dynamics. 

\begin{figure}[tb]
  \centering
  \includegraphics[width=0.95\linewidth]{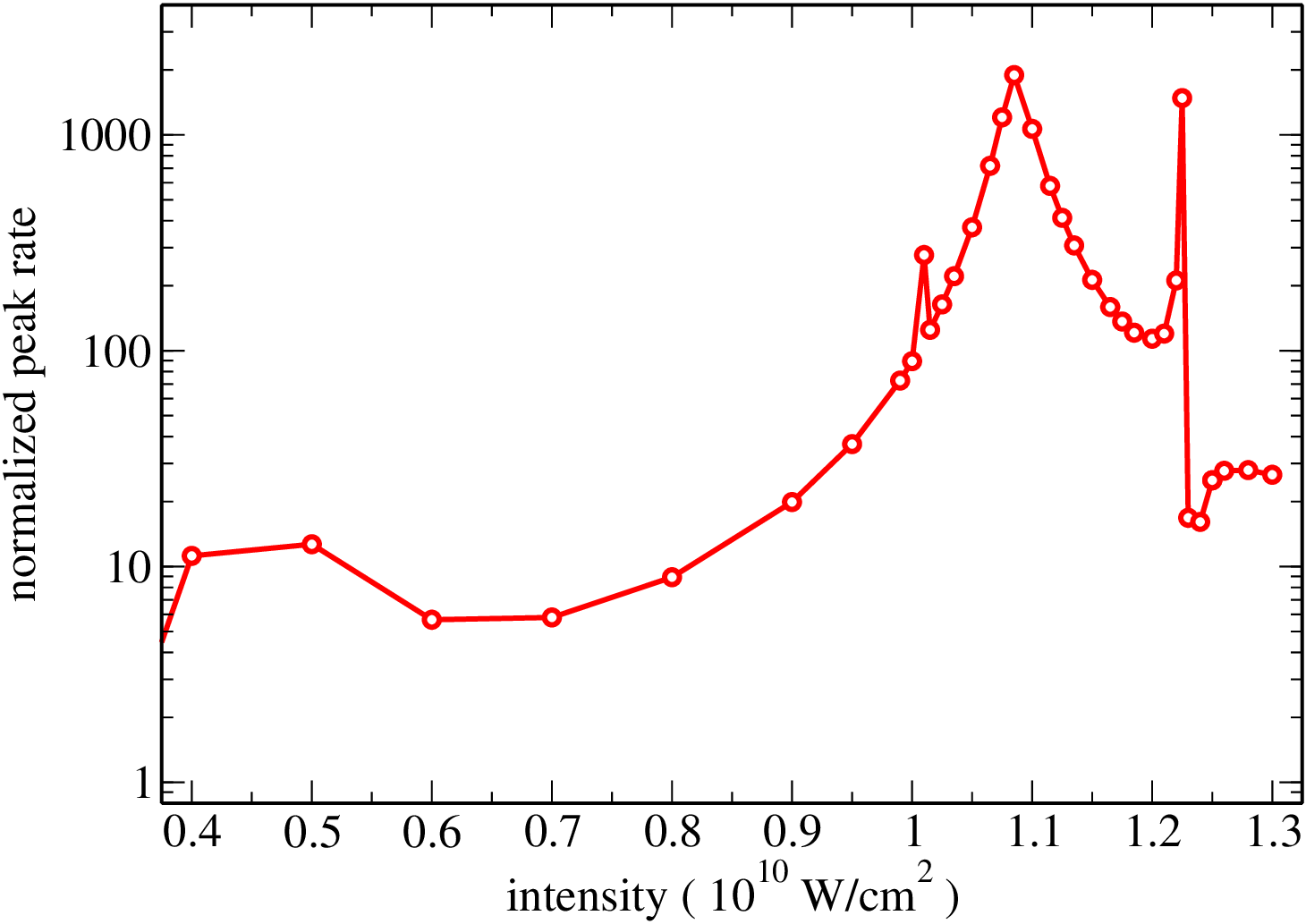}
  \caption{(Color online)
    $\tilde{l}'=1$ peak in the presence of the field 
    normalized with respect to the largest field-free
    peak (with $l'=1$) in the rate coefficients for photoassociation into $\tilde{v}'=-6$
    ($T=1\,\mu$K). Enhancement of the photoassociation rate of about three orders of
    magnitude is found when the position of a  shape resonance comes
    close to the trap temperature.
  }
  \label{fig:enhance}
\end{figure}
The largest increase of the $\tilde{l}'=1$ peak in
Fig.~\ref{fig:PA_Sr2} is found for $I=1.10\times 10^{10}$W/cm$^2$
(red dotted curve). A detailed analysis of the increase as a function
of the non-resonant light intensity is presented in Fig.~\ref{fig:enhance}
which shows  the height of the $\tilde{l}'=1$ peak normalized with
respect to the largest field-free ($l'=1$) peak. 
Most importantly, Fig.~\ref{fig:enhance} demonstrates an
enhancement of the photoassociation rate between two to three orders of magnitude
over a broad range of non-resonant field intensities. 
The narrow peaks at $I=1.01\times 10^{10}\,$W/cm$^2$ and 
$I=1.225\times 10^{10}\,$W/cm$^2$ are due to the $\tilde l=12$ and
$\tilde l=8$ shape resonances, cf. Fig.~\ref{fig:respos}. 
These peaks reflect how the shape resonances pass through the energy
window that corresponds to the atomic cloud's thermal width.
The broad features are attributed to regular scattering states which
are also affected by the non-resonant field.
For sufficiently large $I$, a scattering state is
pushed below the dissociation threshold (without
symmetry restriction, an even parity state is followed by an odd
parity state)~\cite{RosarioNJP09}. Before this happens, the scattering  
wave function is compressed to short internuclear distances, yielding
an increase in the photoassociation rate. This is observed in 
Fig.~\ref{fig:enhance} for intensities around 
$I=1.085\times 10^{10}\,$W/cm$^2$. The corresponding state
becomes bound at $I\approx 1.22\times 10^{10}\,$W/cm$^2$. Also the
peak near $I=5\times 10^{9}\,$W/cm$^2$ is caused by a
state being pushed below the dissociation threshold. However,
the enhancement is much larger for $I=1.085\times
10^{10}\,$W/cm$^2$ since this
state becomes bound inbetween two shape resonances. 

The sharp features of the peak enhancement shown in Fig.~\ref{fig:enhance} 
clearly reflect the $l=0$ rotational weight of the  $\tilde{l}=8$ and
$\tilde{l}=12$ shape resonances. The small peak at $1.01\,$W/cm$^2$ is
connected to the $|c_{l=0}|^2$ weight of the $\tilde{l}=12$ shape
resonance; and the sharp rise and drop of the normalized peak
rate in Fig.~\ref{fig:enhance} at $1.225\,$W/cm$^2$
reflects the behavior of the $|c_{l=0}|^2$ weight for $\tilde{l}=8$ shape
resonance before it becomes bound. The $\tilde l=16$ shape resonance,
also shown in Fig.~\ref{fig:respos} suffers almost no hybridization
of its angular motion, with $|c_{l=16}|^2$ remaining as high as 0.9 for
$I=6.1\times 10^9\,$W/cm$^2$, just before the resonance becomes
bound. Therefore the dipole coupling to $\tilde v'=-6,\tilde l=1$
remains rather small, and no enhancement due to $\tilde l=16$ shape
resonance is observed in Fig.~\ref{fig:enhance}.

In absolute numbers, the maximum peak in Fig.~\ref{fig:enhance}
corresponds to a rate of $3\times 10^{-11}$cm$^3$/s, comparable to the
photoassociation rates of Cs$_2$ molecules in weakly bound levels of the
$0_g^-(P_{3/2})$ excited state~\cite{FiorettiPRL98}.  
The important difference to Cs$_2$ is due to the predominantly $1/R^6$
long-range behavior of the electronically excited state. This allows
almost all of the photoassociated molecules to decay into
\textit{bound} ground state levels and efficient subsequent Raman
transfer to low lying vibrational states. 

\begin{figure}[tb]
  \centering
  \includegraphics[width=0.95\linewidth]{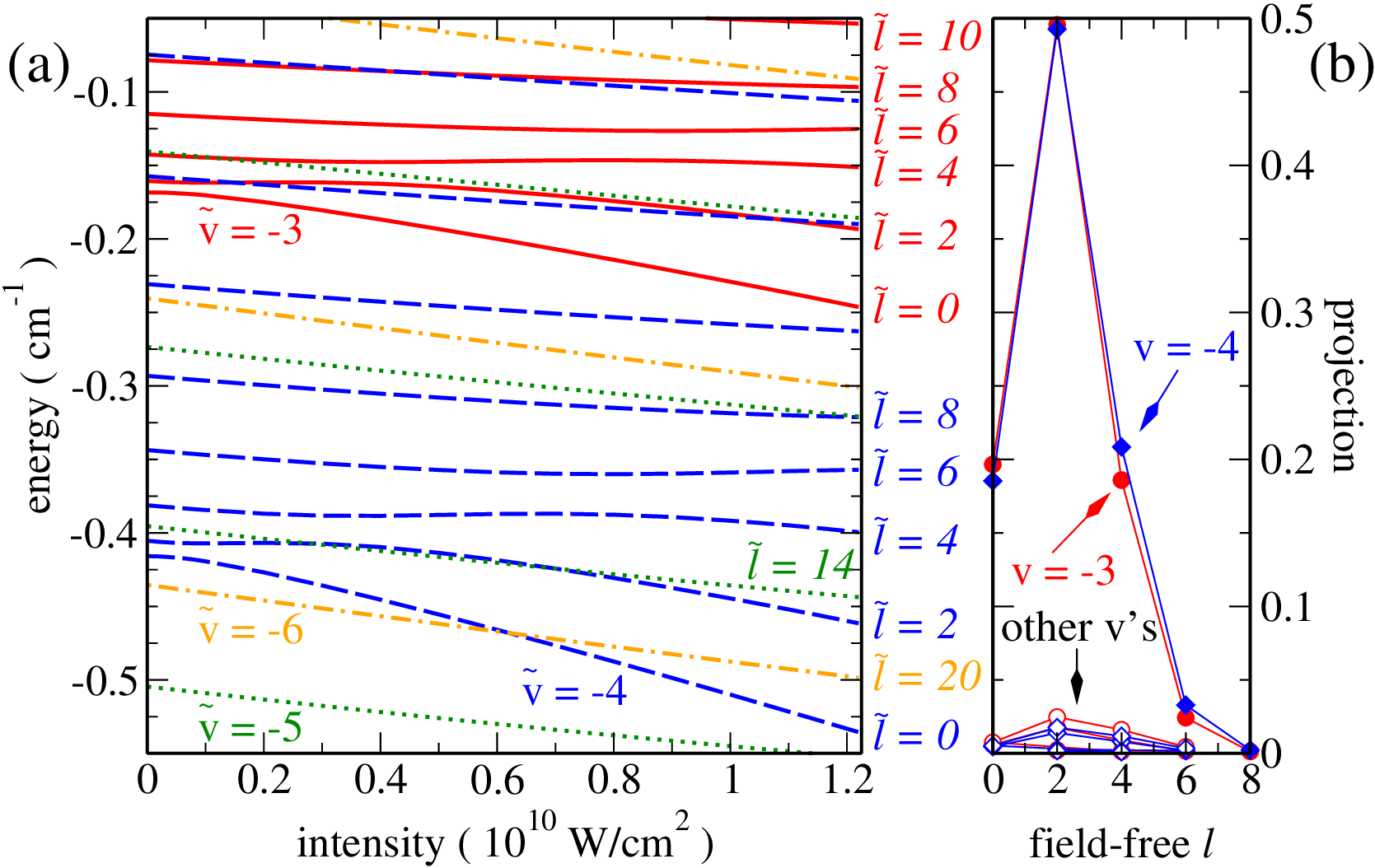}
  \caption{(Color online)
    Energy levels (a) and projections (b) of the hybridized $\tilde
    v''=-3, \tilde l''=0$ (red) and $\tilde 
    v''=-4, \tilde l''=0$ (blue) ground state levels for adiabatic (a) or sudden (b)
    transfer into their field-free counterparts. For a slow switch-off, 
    non-adiabatic transitions are not expected and all molecules are
    obtained with $l''=0$ (a), while for a sudden switch-off, about
    20\% of the molecules have $l''=0$ and about 50\% $l''=2$.
  }
  \label{fig:switchoff}
\end{figure}
The bound states of both ground and excited electronic state  
are characterized by strong alignment in the presence of the
non-resonant field~\cite{FriedrichPRL95}. For example, 
$\langle\cos^2\theta\rangle\gtrsim0.94$ for $I\ge 10^{10}\,$W/cm$^2$.
In the presence of the non-resonant field, the photoassociated
molecules spontaneously decay into hybridized levels of the electronic
ground state. The branching ratios are calculated from the
transition dipole moments~\cite{SkomorowskiKochPRA12} of the field-dressed
rovibrational states. For $I= 1.1 \times 10^{10}\,$W/cm$^2$, 
$\tilde v'=-6,\tilde l'=1$ shows a branching ratio of $31$\% to 
$\tilde v''=-3,\tilde l''=0$ and $43$\% to 
$\tilde v''=-4,\tilde l''=0$ compared to field-free branching ratios of 
$51$\% and $41$\% for $v''=-3, l''=0$ and $l''=2$,
respectively~\footnote{
  The field-free branching ratios differ slightly from those published in
  Ref.~\cite{SkomorowskiKochPRA12} due to the adiabatic approximation
  for the $0_u^+$ state employed here.
}. 
Once molecules in the electronic ground state are
produced, the non-resonant field has to be switched off. 
If this is done adiabatically, the hybridized ground 
state levels $\tilde v''=-3,\tilde l''=0$ and $\tilde v''=-4,\tilde
l''=0$ are directly connected with $v''=-3,l''=0$ and
$v''=-4,l''=0$. A slow switch-off of the non-resonant field transfers
these hybridized levels to their field-free counterparts unless
avoided level
crossings occur. For $\tilde v''=-3,\tilde l''=0$ and $\tilde
v''=-4,\tilde l''=0$, crossings  may be found 
only for very highly excited rotational states from lower vibrational
bands, $\tilde l''\ge 20$, for which the
non-adiabatic couplings with $l''=0$ are small,
cf. Fig.~\ref{fig:switchoff}(a). A 
slow switch-off of the non-resonant field thus produces 
ground state molecules with $v''=-3$ or $v''=-4$ and $l''=0$.
If the molecules after spontaneous decay are subject to loss processes
in the trap, it may not be possible to switch the field off
adiabatically. A sudden switch-off projects the hybridized ground
state levels on a superposition of $m'=0$ field-free rotational levels
of the same vibrational band 
with about 50\% $l'=2$, 20\% $l'=0$, and 20\% $l'=4$ components
for both  $v''=-3$ and $v''=-4$, cf. Fig.~\ref{fig:switchoff}(b). 
Thus, a large part of the molecules
is transferred to ground vibrational levels with $l''=0$ for both
slowly or suddenly switching off the field. The weakly bound ground
state levels show large two-photon transition matrix elements for
efficient Raman transfer to low-lying rovibrational levels of the
electronic ground state~\cite{SkomorowskiKochPRA12}. Almost all of the
molecules produced by photoassociation can thus be transferred to the
rovibrational ground state, in contrast to all previous photoassociation schemes
realized experimentally to date.  

\section{Conclusions}
\label{sec:disc}

We have shown that the photoassociation rate of Sr$_2$ molecules is enhanced 
by three orders of magnitude by applying an additional non-resonant
field. 
This enhancement is due to shape resonances and scattering states
becoming bound, causing a larger number of atom pairs to be 
quasi-trapped at sufficiently short interatomic separations to be
photoassociated. Since the photoassociation rate is limited by the low pair density
at or near the Condon radius~\cite{KochJPhysB06,KochPRA06b}, 
applying strong non-resonant light during photoassociation 
overcomes the main obstacle toward forming larger numbers of
molecules. 
The enhancement of the photoassociation rate is accompanied by strong
hybridization of the angular motion, which we have fully accounted for 
in a rigorous treatment of the coupled rovibrational motion.  

Prior to photoassociation, the non-resonant field needs to be switched on
slowly such that the position of shape resonances and scattering
states is shifted adiabatically to lower
energies. Then the photoassociation laser addresses a thermal cloud that is modified by 
the non-resonant light, producing molecules in an electronically
excited state. As a specific feature of photoassociation near an intercombination
line, the majority of photoassociated molecules 
decays spontaneously into bound ground state levels rather than
re-dissociate~\cite{ZelevinskyPRL06,SkomorowskiKochPRA12}. In the
presence of the non-resonant field, the spontaneous emission involves
hybridized levels. To obtain field-free ground state molecules,
the non-resonant field needs to be
switched off. Ideally, this is done adiabatically, yielding
rotationless molecules in two ground state vibrational
levels. Alternatively, a sudded switch-off leads to a superposition of
ground state molecules with the three lowest rotational quantum
numbers, $0$, $2$ and $4$, occupied.

The required non-resonant field intensities of the order of
$10^{10}\,$W/cm$^2$ can be achieved by employing intra-cavity
beams. Assuming a spot size of the order of 10$\,\mu$m, intra-cavity
powers of up to $10^4\,$W are needed. Such powers can realistically be
produced, for example, by injecting $100\,$W from a single mode
telecom wavelength fiber laser into a cavity with a power
build-up factor of one hundred.  

Enhanced photoassociation rates are a crucial prerequisite to increase the number of
ground state molecules that can be produced optically.
For molecules other than Sr$_2$, our scheme works best for heavy atoms
with large polarizabilities and large scattering lengths. Promising
candidates include, besides other even-isotope alkaline-earth dimers
or Yb$_2$, heteronuclear molecules comprised of one alkali atom and one
alkaline-earth-like atom such as RbSr or RbYb~\cite{MuenchowPCCP11}. 
For these species,
where Feshbach association is either not available or 
likely not to be feasible~\cite{BruePRL12},
non-resonant field control of
shape resonances paves the way to an efficient all-optical
production of ground state molecules. 

\begin{acknowledgments}
  We would like to thank Michael Drewsen, Ronnie Kosloff, and 
  Robert Moszynski for fruitful discussions. Financial support  
  by the Spanish project FIS2011-24540 (MICINN) as well as the
  Grants FQM-2445 and FQM-4643 (Junta de Andaluc\'{\i}a) is
  gratefully acknowledged. RGF belongs to the Andalusian research group
  FQM-207.  
\end{acknowledgments}

\end{document}